\documentclass[twocolumn,epjc3]{svjour3}  
\smartqed
\usepackage[utf8,latin1]{inputenc}
\newcommand{\qm}[1]{``#1''}
\usepackage{amsmath}
\usepackage{amssymb}
\usepackage{tabularx}
\usepackage{colortbl}
\usepackage{graphicx}
\usepackage{dcolumn}
\usepackage{xcolor}
\usepackage{bm}
\usepackage{hyperref}
\hypersetup{colorlinks, linkcolor={red},citecolor={blue},urlcolor={blue}}  
\usepackage[square,numbers]{natbib}
\usepackage{hyperref}
\newcommand\ChangeRT[1]{\noalign{\hrule height #1}}

\def\apj{ApJ }

\def\prd{PRD }
\RequirePackage{graphicx}
\journalname{Eur. Phys. J. C}
\begin{document}

\title{Reconstructing wormhole solutions in curvature based Extended Theories of Gravity}
\titlerunning{Reconstructing wormhole solutions in curvature based Extended Theories of Gravity}      

\author{Vittorio De Falco\thanksref{e1,addr1}
        \and
        Emmanuele Battista\thanksref{e2,addr2,addr3}
        \and
        Salvatore Capozziello\thanksref{e3,addr4,addr5,addr6,addr7}
        \and
        Mariafelicia De Laurentis\thanksref{e4,addr4,addr6,addr8}
}

\thankstext{e1}{e-mail: vittorio.defalco@physics.cz}
\thankstext{e2}{e-mail: emmanuele.battista@kit.edu}
\thankstext{e3}{e-mail: capozziello@unina.it}
\thankstext{e4}{e-mail: mariafelicia.delaurentis@unina.it}

\authorrunning{De Falco V. et al. (2020)}

\institute{Department of Mathematics and Applications \qm{R. Caccioppoli}, University of Naples Federico II, Via Cintia, 80126 Naples, Italy \label{addr1}
           \and
           Institute for Theoretical Physics, Karlsruhe Institute of Technology (KIT), 76128 Karlsruhe, Germany \label{addr2}
           \and
           Institute for Nuclear Physics, Karlsruhe Institute of Technology (KIT), Hermann-von-Helmholtz-Platz 1, 76344 Eggenstein-Leopoldshafen, Germany \label{addr3}
           \and
           Universit\`{a} degli studi di Napoli \qm{Federico II}, Dipartimento di Fisica \qm{Ettore Pancini}, Complesso Universitario di Monte S. Angelo, Via Cintia Edificio 6, I-80126 Napoli, Italy \label{addr4}
           \and
           Scuola Superiore Meridionale, Largo S. Marcellino 10, 80138 Napoli, Italy \label{addr5}
           \and
           Istituto Nazionale di Fisica Nucleare, Sezione di Napoli, Complesso Universitario di Monte S. Angelo, Via Cintia Edificio 6, 80126 Napoli, Italy \label{addr6}
           \and
           Tomsk State Pedagogical University, ul. Kievskaya, 60, 634061
Tomsk, Russia \label{addr7}
           \and
           Lab.Theor.Cosmology,Tomsk State University of Control Systems and Radioelectronics(TUSUR), 634050 Tomsk, Russia \label{addr8}
}

\date{Received: \today / Accepted: }

\maketitle

\begin{abstract}
Static and spherically symmetric wormhole solutions can be reconstructed in the framework of curvature based Extended Theories of Gravity. In particular, extensions of the General Relativity, in metric and curvature formalism give rise to modified gravitational potentials, constituted by the classical Newtonian potential and Yukawa-like corrections, whose parameters can be, in turn, gauged by the observations. Such an approach allows to reconstruct the spacetime out of the wormhole throat considering the  asymptotic flatness as a physical property for the related gravitational field. Such an argument can be applied for a large class of curvature theories characterising the wormholes through the parameters of the potentials. According to this procedure, possible  wormhole solutions could be observationally constrained. On the other hand, stable and traversable wormholes could be a direct probe for this class of Extended Theories of Gravity. 
\end{abstract}

\section{Introduction}
\label{sec:intro}
Wormholes (WHs) are exotic compact objects characterized  by no horizon and singularities, and endowed with a  traversable bridge, called WH neck, connecting two universes or two different regions of the same spacetime \cite{Visser1995}. They have been extensively investigated in the literature, and the papers on such a topic can be distinguished in two categories: (1) finding/constructing solutions in General Relativity (GR) or in extended/ alternative theories of gravity \cite{Visser1989,Barcelo1999,Bohmer2012,Anchordoqui:2000ut,Bahamonde:2016jqq,Capozziello2020}; (2) proposing methods to observationally determine their existence considering reliable fluids or astrophysical probes capable of making these objects stable and traversable structures \cite{Cardoso2016,Konoplya2016,Paul2019,Dai2019,Banerjee2019,Hashimoto2017,Dalui2019,Defalco2020WH}. Regarding the last point, the mentioned strategies are all based on analysis in full and strong  gravitational field regimes. However, at the best of our knowledge, there are no methods to investigate these systems in the weak field limit and then observe their features as realistic gravitational fields. Here we develop a possible strategy to reconstruct WH solutions considering properties of spacetime in the weak field limit of curvature based \emph{Extended Theories of Gravity} \cite{Nojiri2010,Capozziello2011,Clifton2012,Nojiri2017}.
 
In this class of theories, the Ricci curvature scalar $R$ of the Hilbert-Einstein action is replaced by  generic functions of curvature invariants and auxiliary scalar fields. Quantum and cosmological motivations for adopting these GR extensions  are discussed in details in Ref. \cite{FC2011}. A straightforward realization of this approach is done by considering $f(R)$ gravity theories, where GR is just a particular case of a wide class of models (i.e., $f(R)=R$). In this context, also WH solutions can be investigated considering the fact that further degrees of freedom, related to curvature based Extended Theories of Gravity, can give rise to effective perfect fluids acting as sources in their field equations \cite{Zubair2016,Mazharimousavi2016,Mantica,Elizalde2018}. Thanks to these additional geometrical contributions, WHs could be, eventually, stable and traversable depending on the parameters of the belonging theory.

However, WH solutions depend also on the external gravitational field and its asymptotic behaviour, so an \emph{inverse scattering technique} could be employed to reconstruct them from the weak field limit of curvature based Extended Theories of Gravity. It has been shown that from the \emph{Post-Newtonian (PN) expansion} of extended gravity field equations, \emph{Yukawa-like corrections} to the standard Newtonian potential naturally emerge \cite{Arturo,Napolitano2012,Liu2017}. It seems a general feature of several classes of gravitational models, where the particular case is just represented by GR, because such corrections are not present. The parameters of the Yukawa-like corrections can be constrained by different sets of data \cite{DeMartino2018,DeLaurentis2018,Capozziello2020,Capozziello2020FP}, allowing thus to reconstruct reliable WH models in agreement with the observations.
 
\emph{An idea is to describe static and spherically symmetric WHs by Taylor-expanding their metric components in the weak-field limit in order to control how these solutions behave asymptotically. Adopting the weak field limit of curvature based Extended Theories of Gravity, where the parameters are gauged by the observations, and comparing them with the WH expansions, one is able to determine the coefficients of the WH metric and then reconstruct such solutions within different gravity frameworks with the aim to obtain stable and traversable solutions}. This is the central argument on which this paper is based.

An important remark is in order at this point. Theories we are going to take into account are just a particular class of possible extensions of GR. Here we are going to  consider gravity models  based on the Riemann tensor, and the other curvature invariants,  constructed by the Levi-Civita connection of spacetime metric, in other words gravity formulated in the so-called  metric approach.  However, Extended Theories of Gravity can involve also an affine connection independent of the metric, the so-called metric-affine gravity (see e.g. \cite{Olmo}), or a purely affine formulation \cite{Poplawski}. Important sub-classes of metric-affine theories are the Poincar\'e gauge gravity \cite{Poincare}, the modified teleparallel gravity based on the Weitzenb\"ock connection \cite{teleparallel,Hohmann2018}, the modified symmetric teleparallel gravity \cite{Conroy2018}. All these formulations point out that the debate on the fundamental variables describing the gravitational field is still open.

The article is organized as follows: in Sec. \ref{sec:WH} we summarize the properties of static and spherically symmetric WHs; in Sec. \ref{sec:PNE}, the PN expansion in the $f(R)$ gravity framework is discussed as a valid and general paradigm for all curvature based Extended Theories of Gravity; in Sec. \ref{sec:Application} we apply the above-mentioned strategy to constrain the WH solutions through the entries of extended gravity models; finally in Sec. \ref{sec:end} we draw the conclusions. 

\section{Static and spherically symmetric wormhole solutions}
\label{sec:WH}
A static and spherically symmetric WH can be described, in spherical coordinates $(t,r,\theta,\varphi)$ and geometrical units $G=c=1$, by the metric \cite{Morris1988},  
\begin{equation} \label{eq:MTmetric}
\begin{aligned}
&ds^2=-e^{2\chi(r)}dt^2+\frac{dr^2}{1-b(r)/r}+r^2d\Omega^2,
\end{aligned}
\end{equation}
where $d\Omega^2=d\theta^2+\sin^2\theta d\varphi^2$, and $\chi(r)$ and $b(r)$ represent the redshift and shape functions, respectively. Eq.(\ref{eq:MTmetric}) describes a two-parameters family of metrics depending on $\chi(r),b(r)$.  It represent a class of solutions valid both in GR and in extended/alternative theories of gravity, which can be built up to be traversable and stable.  
\begin{figure} [h!]
\centering
\includegraphics[scale=0.47]{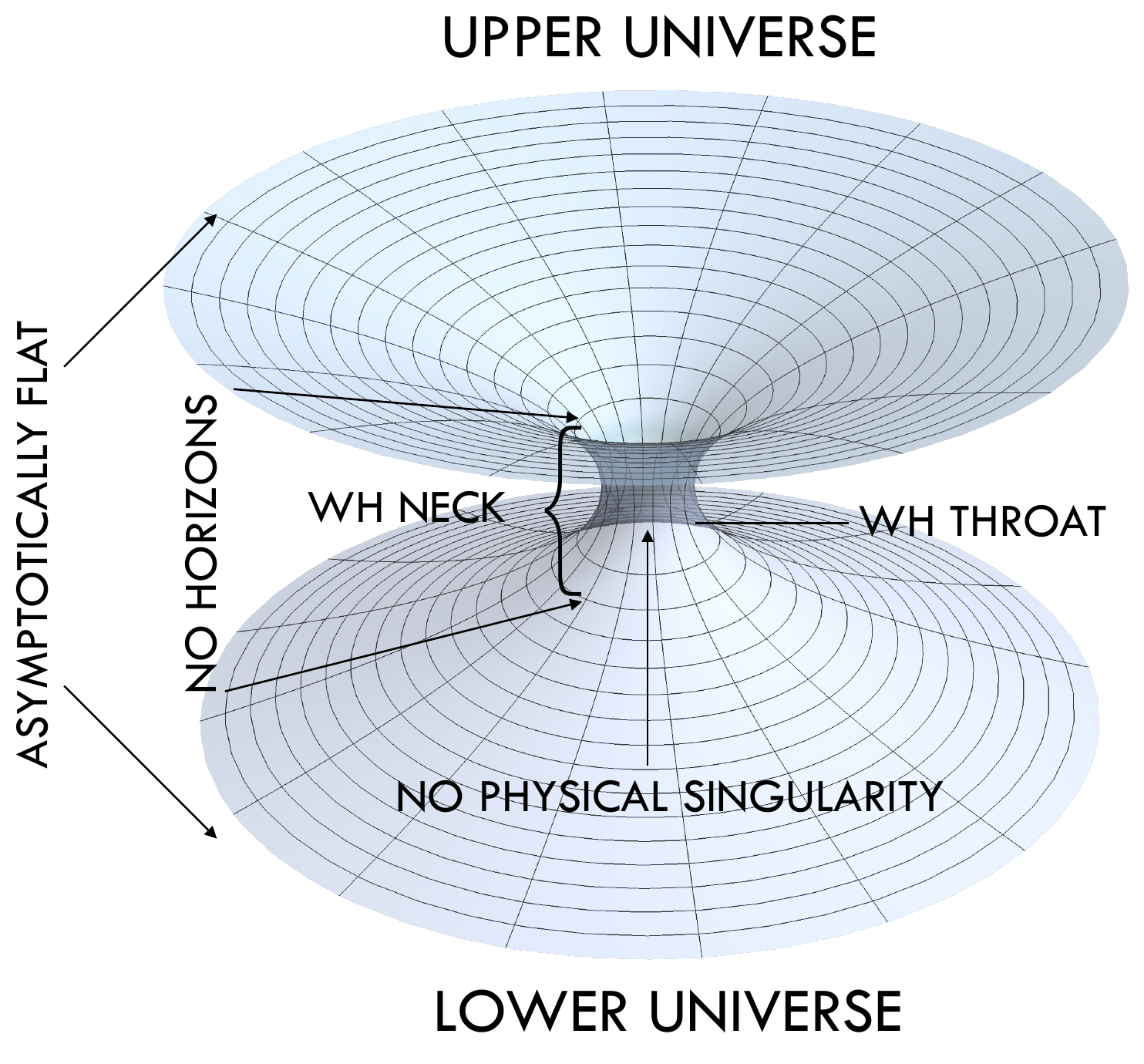}
\caption{Sketch of WH geometry.}
\label{fig:Fig1}
\end{figure}
In Fig. \ref{fig:Fig1} a sketch of the WH geometry is displayed.

Referring to this picture, the significant geometrical proprieties of metric (\ref{eq:MTmetric}) can be summarized  as follows \cite{Morris1988} (see Fig. \ref{fig:Fig1}): (1) the absence of horizons and singularities, which entails that $\chi(r)$ and $b(r)$ are real  smooth functions, and $\chi(r)$ is  everywhere finite; (2) the condition $(1-b(r)/r)\ge0$ allows to define a finite proper radial distance $l$ \cite{Morris1988}; (3) the flaring outward condition \cite{Hochberg19981,Hochberg19982,caplobo1,caplobo2} requires  that $b^{\prime}(r) < b(r)/r$ near and on the throat. It defines the minimum radius such that $r_{\rm min}=b_0$ and $b(r_{\rm min})=b_0$; (4) the asymptotic flatness entails that $b(r)/r\to0$ and $\chi(r)\to0$ for $r\to + \infty$; (5) the WH traversability depends on the underlying theory of gravity. It   can be achieved by considering some form of  exotic matter \cite{Hochberg1997,Bronnikov2013,Garattini2019},  topological defects \cite{Lobo2009,Harko2013,Digrezia2017} or alternative theories of gravity \cite{Capozziello:2012hr}. It is not satisfied in GR for standard perfect fluids; (6)  the mass $M$ is  defined according to the   Arnowitt, Deser, Misner (ADM) formalism. It represents the total mass of the system contained in the whole spacetime \cite{Visser1995}. It is 
\begin{equation} \label{eq:ADMmass}
M\equiv \lim_{r\to+\infty}m(r)=\frac{c^2 b_0}{2G}+4\pi c^2\int^{\infty}_{b_0}\rho(x)x^2 dx.
\end{equation}
By a rapid inspection of the above requirements, in particular properties (3) and (4), it is clear that the external gravitational field and the asymptotic conditions play a major role in defining self-consistent WH solutions. Here we want to investigate a class of alternative theories of gravity (Extended Gravity) which can potentially satisfy the above conditions. The paradigm is that constraining the parameters of gravitational potentials of such models is possible, in principle, to reconstruct WH solutions.  

\section{The Post-Newtonian expansion and the  corrections to the gravitational potential}
\label{sec:PNE}
We consider the external solution of an isolated compact object, where gravity is described within Extended Theories of Gravity. As said above, we mean extensions of GR, where the GR itself is a particular case of a large class of theories. The aim is to reconstruct, in this framework, the features of WH solutions. A straightforward extension of GR is given by the following action 
\begin{equation} \label{eq:Action}
\mathcal{A}=\int f(R) \sqrt{-g}\ {\rm d}^4x,
\end{equation}
where the Lagrangian density $f(R)$ is a generic analytic function of the Ricci curvature scalar $R$ (for $f(R)=R$, GR is restored). Here $g$ is the determinant of the metric tensor $g_{\mu\nu}$, and $\sqrt{-g}\ {\rm d}^4x$ is the invariant volume.

Minimizing the action (i.e., $\delta \mathcal{A}=0$) and computing the integral (\ref{eq:Action}) in the local inertial frame (see Ref. \cite{FC2011}, for details), we obtain the well known fourth-order field equations within the $f(R)$ gravity theories
\begin{equation} \label{eq:FE}
\begin{aligned}
&f'(R)R_{\mu\nu}-\frac{1}{2}f(R)g_{\mu\nu}-f'(R)_{;\mu\nu}\\
&\hspace{1cm}+g_{\mu\nu}\Box f'(R)=\kappa T_{\mu\nu},
\end{aligned}
\end{equation}
with the trace given by
\begin{equation} \label{eq:Trace}
3\Box f'(R)+f'(R)R-2f(R)=\kappa T,
\end{equation}
where $\kappa=8\pi G/c^4$, $f'(R)\equiv df(R)/dR$, $(\cdot)_{;\mu}=\nabla_\mu(\cdot)$ is the covariant derivative, $\Box\equiv g^{\mu\nu}\nabla_\mu\nabla_\nu$ is the curved d'Alembert operator, and $T_{\mu\nu}$ is the stress-energy tensor of a perfect fluid matter. Equations (\ref{eq:FE}) generate a fourth-order dynamics in the metric tensor $g_{\mu\nu}$ \footnote{The field equations (\ref{eq:FE}) can be recast as 
\begin{eqnarray}
G_{\mu\nu}&&=\frac{1}{f'(R)}\left[T^{\rm (curv)}_{\mu\nu}+T_{\mu\nu}\right]\equiv\frac{1}{f'(R)}\left\{f'(R)_{;\mu\nu}-g_{\mu\nu}\Box f'(R)\right.\notag\\
&&\left.\hspace{1.5cm}+g_{\mu\nu}\frac{[f(R)-f'(R)R]}{2}\right\}+\frac{T_{\mu\nu}}{f'(R)},
\end{eqnarray}
where we recover the Einstein tensor $G_{\mu\nu}$ and the higher order terms become part of an effective stress-energy tensor $T^{\rm (curv)}_{\mu\nu}$, which behaves as a perfect fluid, see Ref. \cite{Mantica}, for more details.}.

Since $f(R)$ is an analytic function of $R$, we can expand it in Taylor series around $R=0$ (corresponding to the Minkowski Ricci curvature) obtaining \cite{FC2011}
\begin{equation}
f(R)\approx f_1R+\frac{1}{2}f_2R^2+\dots,\quad f_n=\frac{1}{n!}\frac{d^nf(R)}{dR^n}\Big|_{R=0}.
\end{equation}

Let us consider  a class of static and spherically symmetric metrics described in geometric units $G=c=1$ and spherical coordinates $(t,r,\theta,\varphi)$ as
\begin{equation}
ds^2=-g_{tt}(r)dt^2+g_{rr}(r)dr^2+r^2(d\theta^2+\sin^2\theta d\varphi^2),
\end{equation}
where the unknown functions are $g_{tt}(r),g_{rr}(r)$. Then, we take into account the weak field limit of the metric tensor $g_{\mu\nu}$ with respect to the Minkowski background $\eta_{\mu\nu}$, i.e., $g_{\mu\nu}=\eta_{\mu\nu}+h_{\mu\nu}$ with $1=|\eta_{\mu\nu}|\gg |h_{\mu\nu}|$. 
The PN expansion of the metric components $g_{\mu\nu}(r)$ up to the order $\mathcal{O}(2)$ are given by the following expressions \cite{FC2011}
\begin{eqnarray}
g_{tt}&=&-1 +g_{tt}^{(2)}(r)+g_{tt}^{(4)}(r),\\
g_{rr}&=&-1 +g_{rr}^{(2)}(r),\\
g_{\theta\theta}&=&r^2,\qquad g_{\varphi\varphi}=r^2\sin^2\theta,
\end{eqnarray}
and, at the same order,  the Ricci curvature scalar is
\begin{equation}
R\approx R^{(2)}+R^{(4)}+\dots\ .
\end{equation}
To find the asymptotic expressions of the functions $R^{(2)}$, $g_{tt}^{(2)}$, $g_{rr}^{(2)}(r)$, we need 
three independent equations. Therefore, considering the $tt$ and $rr$ components of Eqs. (\ref{eq:FE}) together with the trace (\ref{eq:Trace}), we respectively have \cite{FC2011}
\begin{eqnarray}
&& g_{tt,rr}^{(2)}+\frac{2}{r}g_{tt,r}^{(2)}-\left[\frac{R^{(2)}(f_1+4f_2)+8f_2R_{,r}^{(2)}}{f_1}\right] =0, \label{eq:FEtt}\\
&& g_{rr,r}^{(2)}-\frac{f_1 r R^{(2)}+8f_2R_{,r}^{(2)}-f_1 r g_{tt,rr}^{(2)}}{2f_1}=0,\label{eq:FErr}\\
&&R_{,rr}^{(2)}+\frac{2}{r}R_{,r}^{(2)}+ \frac{f_1}{6f_2} R^{(2)}=0. \label{eq:FEtra}
\end{eqnarray}
Here we are considering \emph{solutions in vacuum} (i.e., $T_{\mu\nu}=0$), but with the presence of matter the extension of the above solutions can be straightforwardly obtained.

Solving first Eq. (\ref{eq:FEtra}) in terms of $R^{(2)}$, then Eq. (\ref{eq:FEtt}) for $g_{tt}^{(2)}$, and finally Eq. (\ref{eq:FErr}) for $g_{rr}^{(2)}$, we obtain \cite{FC2011}
\begin{eqnarray} 
R^{(2)}&=&\delta_2\frac{e^{-r/L}}{r}+\delta_3\frac{e^{r/L}}{2r}, \label{eq:PN1}\\
g_{tt}^{(2)}&=&\delta_0-\frac{\delta_1}{f_1r}+\delta_2L^2\frac{e^{-r/L}}{3r}+\delta_3L^3\frac{e^{r/L}}{6r}, \label{eq:PN2}\\
g_{rr}^{(2)}&=&-\frac{\delta_1}{f_1r}-\delta_2L^2\left(1+\frac{r}{L}\right)\frac{e^{-r/L}}{3r}\notag\\
&&+\delta_3L^3\left(1-\frac{r}{L}\right)\frac{e^{r/L}}{6r}, \label{eq:PN3}
\end{eqnarray}
where $L=\sqrt{-6f_2/f_1}$, $\delta_0,\delta_1,\delta_2,\delta_3$ are all constants. 

In agreement with whether $L$ is real or complex, we can have different solutions (see Ref. \cite{FC2011}, for details). Assuming  $L$  real, which strictly depends on the sign of $f_1,f_2$, and imposing that at infinity Eqs. (\ref{eq:PN1}) -- (\ref{eq:PN3}) must reduce to the Minkowski metric, we have \begin{eqnarray}
g_{tt}(r)&=&1-\frac{GM}{f_1r}+\frac{\delta_2L^2e^{-r/L}}{3r},\\
g_{rr}(r)&=&1+\frac{GM}{f_1r}+\frac{\delta_2L^2(1+r/L)e^{-r/L}}{3r},\\
R&=&\delta_2\frac{e^{-r/L}}{r}.
\end{eqnarray}
We can combine the constants $f_1$ and $\delta_2$ in one single constant $\delta$ and rewrite the metric components $g_{tt}$ and $g_{rr}$ as gravitational potentials $\Phi$ and $\Psi$, i.e., $g_{tt}=-1+\Phi$ and $g_{rr}=1+\Psi$, which respectively leads to \cite{FC2011}
\begin{eqnarray}
\Phi(r)&=&-\frac{2GM}{rc^2(\delta+1)}(1+\delta e^{-r/L}),\\
\Psi(r)&=&\Phi(r)-\frac{2GM}{rc^2(\delta+1)}\left(\delta\frac{r}{L} e^{-r/L}-2\right),
\end{eqnarray}
where 
\begin{equation}
\delta_2=-\frac{6GM}{L^2}\left(\frac{\delta}{1+\delta}\right),\qquad \delta=-\frac{\delta_2 f_1 L^2}{6GM}.
\end{equation}
Readjusting the parameters in the following way 
\begin{equation}
G\to\frac{2G}{1+\delta},\qquad \alpha\to \delta,\qquad L\to \frac{1}{\lambda},
\end{equation}
the gravitational potential $\Phi$ becomes
\begin{equation} \label{eq:YPOT}
\Phi=-\frac{GM}{r}\left(1+\alpha e^{-\lambda r}\right),
\end{equation}
which is composed by the standard  Newtonian  potential $\Phi_N=-GM/r$ and a Yukawa-like correction $\Phi_Y=\alpha\Phi_N e^{-r/\lambda}$, where $\alpha$ gives the strength of the correction and $\lambda$ is the  length scale over which such potential acts.

The potential $\Psi$ can be also written as
\begin{equation}
\Psi(r)=\Phi(r)+\delta\Phi(r),
\end{equation}
where $\delta\Phi(r)$ is an extra contribution derived from  the above  gravitational potential. Clearly $\Psi(r)\sim\Phi(r)$, as soon as the PN limit of GR is recovered and also it has been showed that in the weak field limit gives very small contributions, implying thus that $\Psi(r)\sim\Phi(r)$ \cite{DeLaurentis2018}.

Similar calculations, developed here for $f(R)$ gravity, can be easily performed for any Extended Theory of Gravity involving higher-order curvature invariants or scalar-tensor terms \cite{Capozziello2011,FC2011}. In Table \ref{tab:Table1}, we report several examples of Extended Theories with the proper corrections and parameterizations of gravitational potentials. 

\section{Constraining wormhole solutions by Extended Theories of Gravity}
\label{sec:Application}
The parameters of Yukawa-like corrections $\Phi_Y$ can be observationally determined considering self-gravitating systems. In  \cite{Capozziello2020FP}, the authors adopted  \emph{the Fundamental Plane of elliptical galaxies} to fulfil this aim. In the specific case of $f(R)$ theories, the range of parameters $f_1,f_2$ (i.e. $\alpha$ and $\lambda$) were fixed by reconstructing models compatible with the Fundamental Plane of elliptical galaxies. In that case, the goal was showing that gravitational corrections could fit dynamics without resorting to the dark matter's existence hypothesis. 

A similar procedure can be adopted here to \qm{reconstruct} WH solutions and their asymptotic behavior in the weak-field limit.
Specifically, we can develop the gravitational potential $\Phi$ in power series of $1/r$, together with the Yukawa term, but avoiding singular expansions at infinity. This procedure will allow thus to restore the prescriptions for physical WHs summarized in Sec.\ref{sec:WH}.

The approach consists in  Taylor expanding the  WH metric components at spatial infinity, obtaining thus two power series of $1/r$ (one for $g_{tt}$ and another one for $g_{rr}$), where the coefficients are functions of the redshift and shape functions, respectively. Then we can match the coefficients of the WH metric with those of the given extended theory of gravity to reconstruct the WH solution. A summary of the models, the modified gravitational potentials, and WH parameters is reported in Table \ref{tab:Table1}.  

\subsection{Asymptotic expansion of the modified Newtonian potential}
\label{sec:PSY}
The expansion of the modified Newtonian potential $\Phi$ of  Eq. (\ref{eq:YPOT}), in power series, means to expand the Yukawa-like potential.  Let us define the variable $x=1/r$, converting thus the limit from $r\to\infty$ to $x\to0$. The function $e^{-\lambda/x}$ has a singular point at $x=0$, which can be extended for continuity, i.e., for $x\to0$ the function $e^{-\lambda/x}\to0$ for all positive values of $\lambda$. If we consider the Taylor expansion of this function at $x=0$, we obtain that the related power series has all coefficients identically equal to zero. This means that, in such a point, the  function $e^{-\lambda/x}$ is approximated by the horizontal straight line $y=0$. If we expand this function in terms of the Laurent series around $x=0$, it gives $\sum_k \frac{(-1)^k}{k!}\lambda^{-k} x^{-k}$. This expansion diverges close to $x=0$, being therefore physically not meaningful for our tasks. 

Let us develop a  feasible approach combining a correct mathematical treatment with an admissible physical interpretation. We Taylor-expand the Yukawa term $e^{-\lambda/x}$ around $x=\varepsilon\ll1$ up to $x^2$, having thus
\begin{equation} \label{eq:YS}
e^{-\frac{\lambda}{x}}=c_1(\varepsilon,\lambda)
+c_2(\varepsilon,\lambda)x+\mathcal{O}\left[(x-\varepsilon)^2\right],
\end{equation}
where
\begin{equation}\label{eq:c12}
c_1(\varepsilon,\lambda)=\frac{e^{-\frac{\lambda }{\varepsilon }} (\varepsilon -\lambda )}{\varepsilon },\qquad c_2(\varepsilon,\lambda)=\frac{\lambda e^{-\frac{\lambda }{\varepsilon }}}{\varepsilon ^2}.
\end{equation}
Equation (\ref{eq:YS}) strictly depends on $\varepsilon$, which, in turn, depends on the sensibility of the chosen observational instruments, $\Theta_{\rm obs}^\pm$\footnote{We refer here to the sensibility of the instrument, $\Theta_{\rm obs}^\pm$, as related to the error of the measurement process (e.g., $\delta r=\pm 0.01$ Kpc). In Sec. \ref{sec:APPL} we show how to calculate it through an example.}. If the measurements are performed at $X=1/R_{\infty}$, with $R_{\infty}$ very far from the WH throat, we can find the value of $\varepsilon$ such that the remainder of the Taylor expansion is included between $\Theta_{\rm obs}^-,\Theta_{\rm obs}^+$, i.e.,
\begin{equation} \label{eq:ineq}
\begin{aligned}
&\left|\mathcal{O}\left[(X-\varepsilon)^2\right]\right|\equiv\left(\frac{\lambda}{2 \varepsilon ^4}  e^{-\frac{\lambda }{\varepsilon }} |2 \varepsilon -\lambda|\right) (X-\varepsilon )^2,\\
&\hspace{1.5cm}\Theta_{\rm obs}^-<\left|\mathcal{O}\left[(X-\varepsilon)^2\right]\right|<\Theta_{\rm obs}^+.
\end{aligned}
\end{equation}
The final Taylor expansion of the $\Phi$ potential is
\begin{equation}
\Phi=\gamma_1x+\gamma_2x^2+\mathcal{O}(x^3),
\end{equation}
where
\begin{equation} \label{eq:gamma}
\begin{aligned}
\frac{\gamma_1}{GM}&=-\left[1+\alpha c_1(\varepsilon,\lambda)\right],\quad
\frac{\gamma_2}{GM}=-\alpha c_2(\varepsilon,\lambda).    
\end{aligned}
\end{equation}
With these considerations in mind, let us discuss the asymptotic expansion of the WH metric.

\subsection{Asymptotic expansion of the wormhole metric}
\label{sec:PSWH}
Let us Taylor-expand the $g_{tt}$ and $g_{rr}$ components of the WH metric (\ref{eq:MTmetric}) around $x=0$, which contain the unknown functions $\chi(x)$ and $b(x)$, having thus\footnote{We follow the same approach used to determine the asymptotic expansion of the Schwarzschild metric. For example, the radial component $g_{rr}(r)=1/(1-2M/r)$ can be correctly expanded at infinity by using the variable $x=1/r$. We have the function $g_{rr}(x)=1/(1-2Mx)$, and we can expand it in Taylor series around $x=0$, having $g_{rr}(x)=1 + 2 M x + 4 M^2 x^2+\mathcal{O}(x^3)$.}
\begin{eqnarray}
g_{tt}&\equiv& e^{-2\chi(x)}=-1+\alpha_1x+\alpha_2x^2+\mathcal{O}(x^3),\\
g_{rr}&\equiv&\frac{1}{1-b(x)x}=1+\beta_1x+\beta_2x^2+\mathcal{O}\left(x^3\right).
\end{eqnarray}
where
\begin{eqnarray}
&&\alpha_1=-2\chi'(0),\qquad \alpha_2=-2\chi'(0)^2-\chi''(0), \label{eq:alpha}\\
&& \beta_1=b(0),\qquad\quad\ \ \beta_2=b(0)^2+b'(0).\label{eq:beta}
\end{eqnarray}
From these equations, we can easily write the unknown functions and their derivatives evaluated in $x=0$ in terms of the coefficients of their expansions, i.e.,
\begin{eqnarray}
&&\chi'(0)=-\frac{\alpha_1}{2},\qquad \chi''(0)=\alpha_1-\alpha_2,\label{eq:chi}\\
&&b(0)=\beta_1,\qquad\quad\ \ b'(0)=\beta_2-\beta_1^2,\label{eq:b}
\end{eqnarray}
with
\begin{equation}
\begin{aligned}
&\chi'(0)=\left.\frac{d\chi(x)}{dx}\right|_{x=0},\quad \chi''(0)=\left.\frac{d^2\chi(x)}{dx^2}\right|_{x=0},\\ &b(0)=\left.b(x)\right|_{x=0},\qquad \ \ b'(0)=\left.\frac{db(x)}{dx}\right|_{x=0}.
\end{aligned}
\end{equation}

The redshift and shape functions are therefore
\begin{eqnarray}
\chi(x)&=&\chi'(0)x+\frac{\chi''(0)}{2}x^2,\label{eq:chi2}\\ 
b(x)&=&b(0)+b'(0)x.\label{eq:b2}
\end{eqnarray}
Now, we can explicitly determine the parameters (\ref{eq:chi}) -- (\ref{eq:b}) to reconstruct the WH metric (\ref{eq:MTmetric}) by matching them with those of the Newtonian potential and the Yukawa potential (\ref{eq:gamma}) within a precise gravity framework and, eventually, benchmark them with the observational data.

\subsection{Reconstructing wormhole solutions}
\label{sec:MWH}
\begin{figure*} [t!]
    \centering
    \hbox{
    \includegraphics[scale=0.29]{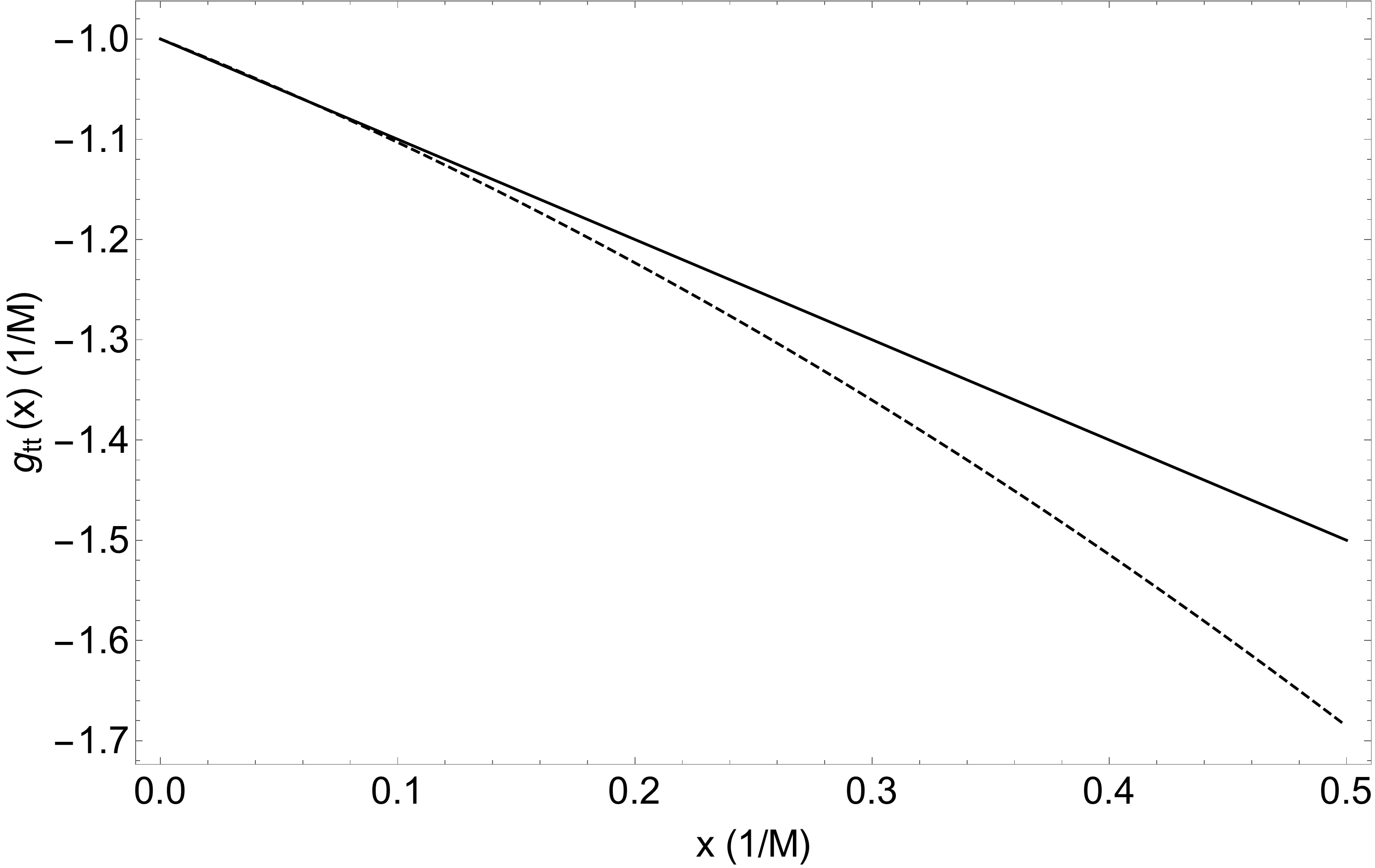}
    \hspace{0.5cm}
    \includegraphics[scale=0.29]{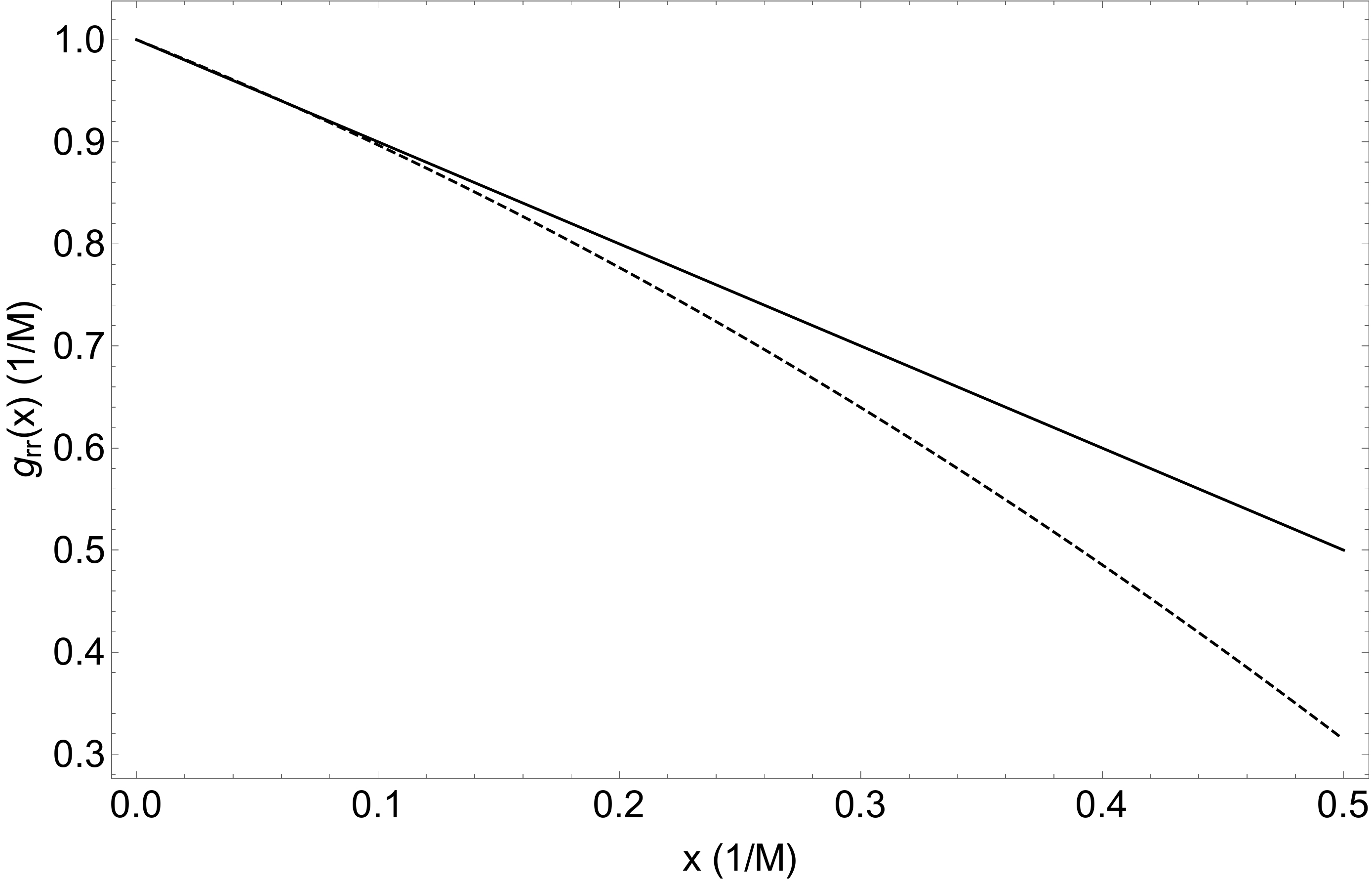}}
    \vspace{0.5cm}
    \hbox{
    \includegraphics[scale=0.29]{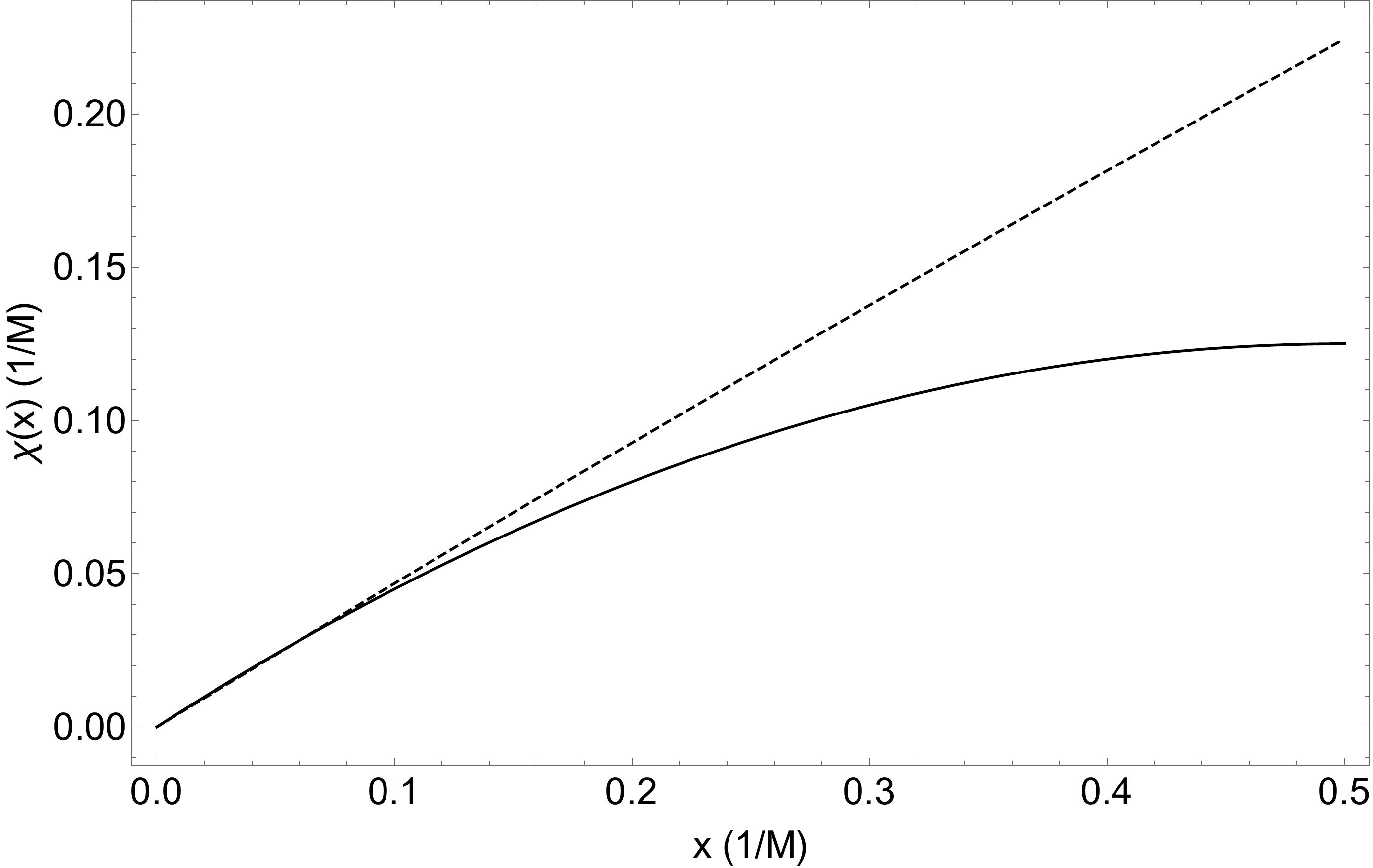}
    \hspace{0.5cm}
    \includegraphics[scale=0.29]{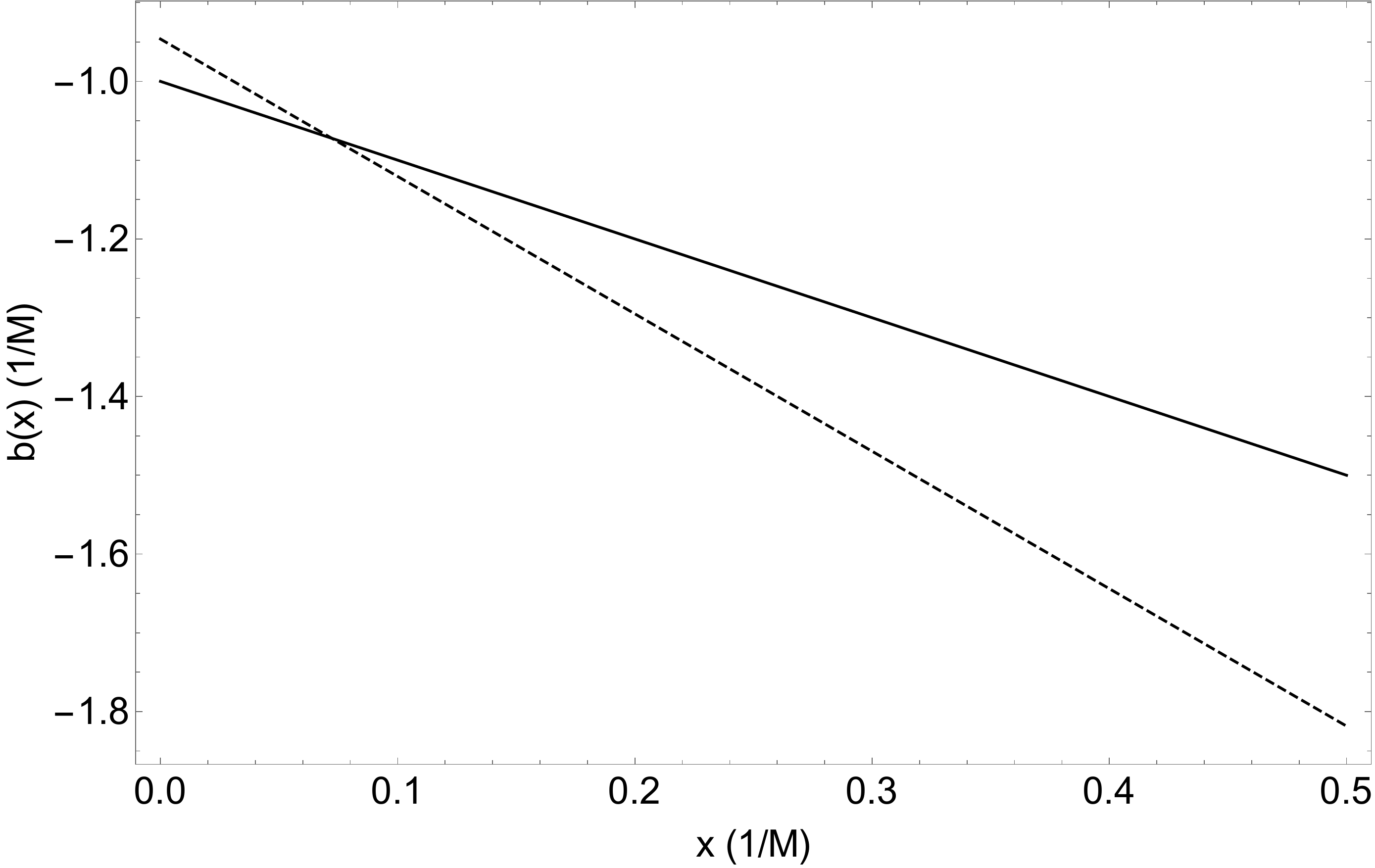}}
    \caption{WH metric time $g_{tt}(x)$ and radial $g_{rr}(x)$ components, and redshift $\chi(x)$ and shape $b(x)$ functions plotted in terms of the inverse of the radial distance $x$ for $f(R)$ (continuous lines) and $R+\alpha_0 R^2+\alpha_1 R\Box R$ (dashed lines) theories. The WH solution has been reconstructed by considering the globular cluster \texttt{NGC 4649}, having an effective radius of $R_\infty=10.00$ Kpc, $\varepsilon=0.111828\ \mbox{Kpc}^{-1}$ for the expansion of the Yuakawa potential, and the best fit values for the modified potential (\ref{eq:YPOT}) are $\lambda=1\ \mbox{Kpc}^{-1}$, and $\alpha=0.01$ for $f(R)$ theories, and $d_0=0.56$, $d_1=-0.12$, $L_0=3.77$, $L_1=1.95$ for $R+\alpha_0 R^2+\alpha_1 R\Box R$ theories.}
    \label{fig:Fig2}
\end{figure*}

We consider the modified theories of gravity reported in Table I of Ref. \cite{Capozziello2020FP}. In Table \ref{tab:Table1} we summarise these theories of gravity together with related modified Newtonian potentials, and coefficients $\gamma_1,\gamma_2$ of their asymptotic expansions given in  Eqs. (\ref{eq:gamma}). By considering Eqs. (\ref{eq:alpha}) -- (\ref{eq:beta}), we can reconstruct the WH expansion by matching the coefficients of the two power series, i.e.,
\begin{equation}
\alpha_1=\gamma_1,\quad \alpha_2=\gamma_2,\quad \beta_1=\gamma_1,\quad \beta_2=\gamma_2.     
\end{equation}
Substituting them in Eqs. (\ref{eq:chi}) -- (\ref{eq:b}), we reconstruct the WH redshift and shape functions in Eqs. (\ref{eq:chi2}) and  (\ref{eq:b2}).

\subsubsection{Application of the method to specific examples}
\label{sec:APPL}
As an example, let us consider the globular cluster \texttt{NGC 4649}, also known as \texttt{M60}, because it is located relatively close to the Earth ($\sim60$ millions of light years), it is the third-brightest giant elliptical galaxy of the Virgo cluster of galaxies, and it has several other advantageous proprieties, which permit to be easily detected and studied (see Ref. \cite{DeBruyne2001}, for more details). It has an effective radius $R_\infty=10.00$ Kpc \cite{DeBruyne2001}, and it hosts at its center a supermassive black hole of mass $(4.5 \pm 1.0) \times 10^9\ M_\odot$ (see Ref. \cite{Shen2010}, for details), which we clearly assume to be a WH, in lack of data on such exotic compact objects.

Following the procedure in \cite{Capozziello2020FP}  to constrain the modified Newtonian potential (\ref{eq:YPOT}), we use their best fit values, i.e., $R_\infty\cdot\lambda=10$ or $\lambda=1\ \mbox{Kpc}^{-1}$, and $\alpha=0.01$ (see Fig. 7 in Ref. \cite{Capozziello2020FP}, for details). To determine the value of $\varepsilon$, we know that the error in the measurement of the effective radius (or also distance from the compact object) is $\delta R_\infty=\pm0.01$ Kpc. Therefore the detection sensibility is $\Theta_{\rm obs}^\pm\equiv e^{-\lambda (R_\infty\pm\delta R_\infty)}$, where $\Theta_{\rm obs}^-=4.5\times 10^{-5}$ and $\Theta_{\rm obs}^+=4.6\times 10^{-5}$. Adopting the inequality (\ref{eq:ineq}) with $X=1/R_\infty$, we find that $0.111781\ \mbox{Kpc}^{-1}\lesssim\varepsilon\lesssim0.111876\ \mbox{Kpc}^{-1}$, and we choose $\varepsilon=0.111828\ \mbox{Kpc}^{-1}$. Now, we can calculate the coefficients $\gamma_1,\gamma_2$ by selecting two extended theories of gravity from Table \ref{tab:Table1}. We consider for simplicity $f(R)$ and $R+\alpha_0 R^2+\alpha_1 R\Box R$, choosing $\alpha_0=-3$ and $\alpha_1=-9$\footnote{Looking at Table \ref{tab:Table1}, we see that in order to have $d_{0,1}$ and $L_{0,1}$ real, it is easy to impose the following inequalities on the theory parameters $\alpha_0,\alpha_1$, namely $\alpha_1\leq0$ and $\alpha_0\leq -\sqrt{-2a_1/3}$. }. In such hypothesis we obtain the following results:
\begin{itemize}
    \item $f(R)$ theory: the expansion coefficients (\ref{eq:gamma}) for $\alpha=0.01$, $\lambda=1\ \mbox{Kpc}^{-1}$  are $\gamma_1=-1$,
$\gamma_2=-10^{-4}$;
     \item $R+\alpha_0 R^2+\alpha_1 R\Box R$ theory: the expansion coefficients (\ref{eq:gamma}) for $d_0=0.56$, $d_1=-0.12$, $L_0=3.77$, $L_1=1.95$ are $\gamma_1=-0.95$, 
$\gamma_2=-0.85$.
\end{itemize}

In Fig. \ref{fig:Fig2}, we provide the WH reconstruction by showing the behaviours of $g_{tt}(x),g_{rr}(x),\chi(x),b(x)$, which should be considered valid only for values very close to $x=0$, although they have been all plotted up to $r=2M$. This expansion can be easily extended to a generic order $n$, giving more precise results. In addition, such a strategy can be complemented with other techniques in full and strong gravitational field regimes to accurately infer more information on the WH solutions.   

\renewcommand{\arraystretch}{1.8}
\begin{table*}[t!]
\begin{center}
\caption{\label{tab:Table1} Summary of different theories of gravity and related modified Newtonian potentials (see Table I in Ref. \cite{Capozziello2020FP}), and the first two coefficients of the asymptotic expansion of the $\Phi$ potential, see Eq. (\ref{eq:gamma}). The lightgray cells refer to  GR.}	
\normalsize
\vspace{0.3cm}
\scalebox{0.8}{
\begin{tabular}{|@{} c @{}|@{} l @{}|@{} c @{}|@{} c @{}|} 
\ChangeRT{1pt}
\ \ {\bf Theory of gravity}\ \ &\ \ {\bf Modified Newtonian potential}\ \ &\ \ $\boldsymbol{\gamma_1}$\ \ &\ \ $\boldsymbol{\gamma_2}$ \ \ \\
\hline
\rowcolor{lightgray} R & $\Phi_N=-GM x$ & $-GM$ & $0$ \\
\hline
$f(R)$ & $\Phi=\Phi_N\left(1+\alpha e^{-\frac{\lambda}{x}}\right)$ & $-GM[1+\alpha c_1(\varepsilon,\lambda)]$ & $-GM\alpha c_2(\varepsilon,\lambda)$ \\
\hline
$R+\alpha_0 R^2+\alpha_1 R\Box R$ & $\Phi=\Phi_N\left(1+d_0e^{\frac{-1}{L_0x}}+d_1e^{\frac{-1}{L_1x}}\right)$ & $-GM[1+d_0c_1(\varepsilon,1/L_0)$ & $-GM[d_0c_2(\varepsilon,1/L_0)$ \\
\cline{2-2}
& $d_{0,1}=\frac{1}{6}\mp\frac{a_0}{2\sqrt{9a_0^2+6a_1}},$ & $+d_1c_1(\varepsilon,1/L_1)]$ & $+d_1c_2(\varepsilon,1/L_1)]$\\
& $L_{0,1}=\sqrt{-3a_0\pm\sqrt{9a_0^2+6a_1}}$ & &\\
\hline
$R+\sum_{k=0}^p\alpha_k R\Box^k R$ & $\Phi=\Phi_N\left(1+\Sigma_{k=0}^{p}d_k e^{\frac{-1}{L_kx}}\right)$ & $-GM\left[1+\sum_{k=0}^pd_kc_1(\varepsilon,1/L_k)\right]$ & $-GM\sum_{k=0}^pd_kc_2(\varepsilon,1/L_k)$ \\
\cline{2-2}
 & $d_k$ and $L_K$ are functions of $a_k$, & &\\
 & see Ref. \cite{Capozziello2020FP}, for more details & &\\
\hline
$f(R,R_{\alpha\beta}R^{\alpha\beta})$ & $\Phi=\Phi_N\left[1+\frac{1}{3}\,e^{\frac{-m_R}{x} }-\frac{4}{3}\,e^{\frac{-m_Y}{x}}\right]$ & $-GM\left[1+\frac{c_1(\varepsilon,m_R)}{3}-\frac{4c_1(\varepsilon,m_Y)}{3}\right]$ & $-GM\left[\frac{c_2(\varepsilon,m_R)}{3}-\frac{4c_2(\varepsilon,m_Y)}{3}\right]$ \\
\cline{2-2}
& $Y=R_{\mu\nu}R^{\mu\nu},\quad m_Y^2=\frac{1}{f_Y(0)},$ & &\\
& $m_R^2=-\frac{1}{3f_{RR}(0)+2f_Y(0)}$ & &\\
\hline
$f(R,R^2-4R_{\mu\nu}R^{\mu\nu}+R_{\alpha\beta\mu\nu}R^{\alpha\beta\mu\nu})$ & $\Phi=\Phi_N\left[1+\frac{1}{3}\,e^{\frac{-m_1}{x} }-\frac{4}{3}\,e^{\frac{-m_2}{x}}\right]$ & $-GM\left[1+\frac{c_1(\varepsilon,m_1)}{3}-\frac{4c_1(\varepsilon,m_2)}{3}\right]$ & $-GM\left[\frac{c_2(\varepsilon,m_1)}{3}-\frac{4c_2(\varepsilon,m_2)}{3}\right]$ \\
\cline{2-2}
 & $Z=R_{\alpha\beta\mu\nu}R^{\alpha\beta\mu\nu},\quad  m_2^2=\frac{1}{f_Y(0)+4f_Z(0)},$ & &\\
 & $m_1^2=-\frac{1}{3f_{RR}(0)+2f_{YY}(0)+2f_Z(0)}$ & &\\
\hline
$f(R,\phi)+\omega(\phi)\phi_{;\alpha}\phi^{;\alpha}$ & $\Phi=\Phi_N\left[1+g(\xi,\eta)\,e^{\frac{-m_R\tilde{k}_R}{x}}\right.$ & $-GM\left[1+g(\varepsilon,\eta)c_1(\varepsilon,m_R\tilde{k}_R)\right.$ & $-GM\left[g(\varepsilon,\eta)c_2(\varepsilon,m_R\tilde{k}_R)\right.$ \\
 & $\left.+\left(\frac{1}{3}-g(\xi,\eta)\right)\,e^{\frac{-m_R\tilde{k}_\phi}{x}}\right]$ & $\left.+\left(\frac{1}{3}-g(\varepsilon,\eta)\right)c_1(\varepsilon,m_R\tilde{k}_\phi)\right]$ & $\left.+\left(\frac{1}{3}-g(\varepsilon,\eta)\right)c_2(\varepsilon,m_R\tilde{k}_\phi)\right]$\\
\cline{2-2} 
 & $\eta=\frac{m_\phi}{m_R},\quad m_R^2=-\frac{1}{3f_{RR}(0,\phi_0)}$ & &\\
 & $m_\phi^2=-\frac{f_{\phi\phi}(0,\phi_0)}{2\omega(\phi_0)},\quad \xi=\frac{3{f_{R\phi}(0,\phi_0)}^2}{2\omega(\phi_0)},$ & &\\
 & $ g(\xi,\eta)=\frac{1-\eta^2+\xi+\sqrt{\eta^4+(\xi-1)^2-2\eta^2(\xi+1)}}{6\sqrt{\eta^4+(\xi-1)^2-2\eta^2(\xi+1)}}$ & &\\
 & $\tilde{k}_{R,\phi}^2=\frac{1-\xi+\eta^2\pm\sqrt{(1-\xi+\eta^2)^2-4\eta^2}}{2}$ & &\\
\hline
$f(R,R_{\alpha\beta}R^{\alpha\beta},\phi)+\omega(\phi)\phi_{;\alpha}\phi^{;\alpha}$ & $\Phi=\Phi_N\left[1+g(\xi,\eta)\,e^{\frac{-m_R\tilde{k}_R}{x}}\right.$ & $-GM\left[1+g(\varepsilon,\eta)c_1(\varepsilon,m_R\tilde{k}_R)\right.$ & $-GM\left[g(\varepsilon,\eta)c_2(\varepsilon,m_R\tilde{k}_R)\right.$ \\
& $\left.+\left(\frac{1}{3}-g(\xi,\eta)\right)\,e^{\frac{-m_R\tilde{k}_\phi}{x}}-\frac{4}{3}\,e^{\frac{-m_Y}{x}}\right]$ & $+\left(\frac{1}{3}-g(\varepsilon,\eta)\right)c_1(\varepsilon,m_R\tilde{k}_\phi)$ & $+\left(\frac{1}{3}-g(\varepsilon,\eta)\right)c_2(\varepsilon,m_R\tilde{k}_\phi)$\\
\cline{2-2} 
 & $\eta=\frac{m_\phi}{m_R},\quad m_R^2=-\frac{1}{3f_{RR}(0,0,\phi_0)+2f_Y(0,0,\phi_0)},$ & $\left.-\frac{4}{3}c_1(\varepsilon,m_Y)\right]$ & $\left.-\frac{4}{3}c_2(\varepsilon,m_Y)\right]$\\
 & $m_Y^2=\frac{1}{f_Y(0,0,\phi_0)},\quad
 m_\phi^2=-\frac{f_{\phi\phi}(0,0,\phi_0)}{2\omega(\phi_0)}$ & &\\
 & $\xi=\frac{3{f_{R\phi}(0,0,\phi_0)}^2}{2\omega(\phi_0)},$ & &\\
 & $g(\xi,\eta)=\frac{1-\eta^2+\xi+\sqrt{\eta^4+(\xi-1)^2-2\eta^2(\xi+1)}}{6\sqrt{\eta^4+(\xi-1)^2-2\eta^2(\xi+1)}}$ & &\\
 & $\tilde{k}_{R,\phi}^2=\frac{1-\xi+\eta^2\pm\sqrt{(1-\xi+\eta^2)^2-4\eta^2}}{2}$ & &\\
\ChangeRT{1pt}
\end{tabular}}
\end{center}
\end{table*}

\section{Discussion and Conclusions}
\label{sec:end}
We have developed a strategy to reconstruct WH solutions through Extended Theories of Gravity (tuned by fitting the observational data) in the weak gravitational field limit.  The method is based on a model independent framework employing a family of static and spherically symmetric metrics (\ref{eq:MTmetric}), depending upon the two unknown $\chi(r)$ (redshft) and $b(r)$ (shape) functions, see Sec. \ref{sec:WH}.

In the weak field limit, GR simply reduces to the Newtonian theory, but the observations on the rotational curves and on mass-to-light rations of several galaxies showed a clear departure from the classical description. Therefore, to solve such an issue, modified theories of gravity have been proposed, whose true nature can be reconstructed only by the fit of the data. Such theories solve this observational puzzle by adding to the Newtonian potential a Yukawa-like correction (\ref{eq:YPOT}), whose parameters can be gauged by the data, see Sec. \ref{sec:PNE}

Our method consists in first Taylor-expanding the modified Newtonian potential (\ref{eq:YPOT}) around $x\equiv 1/r=0$ (or in weak field limit), where the most delicate part is related to the expansion of the Yukawa-like correction, which behaves singularly at $x=0$. However, we have developed a new procedure, based on Taylor expanding such a function around $x=\varepsilon\ll1$, and determining the value of this parameter through the current sensibility of the observational instruments, see Sec. \ref{sec:PSWH}. Then, we Taylor expand also the WH $g_{tt},g_{rr}$ metric components of Eq. (\ref{eq:MTmetric}) around $x=0$, see Sec. \ref{sec:PSWH}. Finally, we match the coefficients of the two expressions to reconstruct the WH solution from the modified theories of gravity, see Sec. \ref{sec:MWH}. In Table \ref{tab:Table1} we show different theories of gravity with the related modified Newtonian potentials and coefficients $\gamma_1,\gamma_2$ of their asymptotic expansion. As an example of our strategy, we show in Fig. \ref{fig:Fig2} the asymptotic behaviour of the WH metric components, redshift, and shape functions for two different modified theories of gravity, see Sec. \ref{sec:APPL} for more details.

The considerations developed in this paper are useful, in general,  to build up a stable and traversable WH solution within Extended Gravity, because the above procedure permits to impose conditions on the WH behaviour in the weak field limit ruled by the given theory of gravity. Depending on the WH model, this last feature can have also important consequences for selecting the class of (geometric or matter) fluids used to make the WH stable and traversable and of course for the related energy conditions. Taking into account these constraints, it would be possible to rule out  solutions not in agreement with the aforementioned requests and restrict thus the class of admissible and physical WH solutions.

In summary, we are developing, in this paper and in others (see e.g. \cite{Defalco2020WH}) a \emph{model-independent approach}, which can be advantageous because it allows: (1) to search for WH observational existence; (2) to reduce the set of WH solutions capable of  being compared to observational data; (3) to impose constraints on extended or modified theories of gravity; (4) to adapt such strategies for investigating also black holes and other compact objects. In a future paper, we aim at extending this strategy for rotating and axially symmetric WH metrics.

\section*{Acknowledgements}
V.D.F. thanks Gruppo Nazionale di Fisica Matematica of Istituto Nazionale di Alta Matematica (INDAM) for support. S.C. and M.D.L. acknowledge the support of Istituto Nazionale di Fisica Nucleare (INFN) sez. di Napoli, Iniziative Specifiche MOONLIGHT2,  QGSKY, and  TEONGRAV. The authors thank the anonymous referees for the useful comments.

\end{document}